\documentclass[aps,pre,twocolumn,superscriptaddress,showpacs]{revtex4}

\usepackage{graphicx}

\begin{document}

\title{Granular circulation in a cylindrical pan: simulations of reversing
  radial and tangential flows}

\author{Oleh Baran}
\email[]{oleh.baran@njit.edu}
\altaffiliation[Currently at ]{ExxonMobil Research and Engineering,
1545 Route 22 East, Annandale, NJ, USA, 08801.}
\affiliation{Department of Applied Mathematics, University of
Western Ontario, London ON, N6A 5B9, Canada}

\author{John J. Drozd}
\affiliation{Department of Applied Mathematics, University of
Western Ontario, London ON, N6A 5B9, Canada}

\author{Robert J. Martinuzzi} 
\affiliation{Department of Mechanical and Materials Engineering,
  University of Western Ontario, London ON, N6A 5B9, Canada}
\affiliation{Department of Mechanical and Manufacturing Engineering,
  University of Calgary, Calgary AB, T2N 1N4, Canada}

\author{Peter H. Poole}
\affiliation{Department of Applied Mathematics,
University of Western Ontario, London ON, N6A 5B9, Canada}
\affiliation{Department of Physics, St. Francis Xavier University,
Antigonish NS, B2G 2W5, Canada} 

\date{\today}

\begin{abstract}
  Granular flows due to simultaneous vertical and horizontal
  excitations of a flat-bottomed cylindrical pan are investigated
  using event-driven molecular dynamics simulations.  In agreement
  with recent experimental results, we observe a transition from a
  solid-like state, to a fluidized state in which circulatory flow
  occurs simultaneously in the radial and tangential directions.  By
  going beyond the range of conditions explored experimentally, we
  find that each of these circulations reverse their direction as a
  function of the control parameters of the motion.  We numerically
  evaluate the dynamical phase diagram for this system and show, using
  a simple model, that the solid-fluid transition can be understood in
  terms of a critical value of the radial acceleration of the pan
  bottom; and that the circulation reversals are controlled by the
  phase shift relating the horizontal and vertical components of the
  vibrations.  We also discuss the crucial role played by the geometry
  of the boundary conditions, and point out a relationship of the
  circulation observed here and the flows generated in vibratory
  conveyors.
\end{abstract}

\maketitle

\section{Introduction}

In a recent experimental work, Sistla, et al.~\cite{sistla02} observed
a novel granular circulation occurring in a flat-bottomed cylindrical
container (or ``pan'') subjected to coupled horizontal and vertical
vibrations.  The apparatus studied was a commonly available industrial
device used mainly for sieving and polishing of granular particles.
Of particular interest was the appearance of a ``toroidal''
circulation, in which granular flow occurred simultaneously in the
tangential direction (i.e. a circulation in the horizontal plane about
the vertical axis), as well as radially (i.e. particles flowed outward
along the bottom surface and then back to the center of the pan along
the top surface of the granular bed).  In this mode, the trajectory of
a single particle would therefore approximate a helix on the surface
of a torus.  Among several other phenomena, it was noted that the
direction of the tangential flow was, under most of the conditions
explored, opposite to that of the orbital motion of the center of the
pan.  However, under some conditions, the tangential flow would
reverse direction, and move in the same direction as the pan center.
The authors noted that the phase shift relating the vertical and
horizontal vibrations seemed to control this reversal.  Whatever their
origin, these phenomena indicate a non-trivial interplay of the
horizontal and vertical vibrations, and their interaction with the
granular bed.  Since the bottom surface of the pan in the apparatus
was a rough wire screen, the authors of Ref.~\cite{sistla02} suggested
that entrainment of particles to the motion of the bottom surface
might play an important role in establishing the direction of the
observed granular circulations.

Motivated by the complexity of phenomena observed in
Ref.~\cite{sistla02}, we present here a computer simulation study of a
granular matter system enclosed in the same container geometry as in
Ref.~\cite{sistla02}, and subjected to a similar form of horizontal
and vertical vibration.  As we will show below, we are able to
reproduce the major dynamical modes of the granular bed observed
experimentally. We also explore the dynamical behavior beyond the
range studied experimentally, and discover an entire family of
circulation reversal transitions, affecting both the tangential and
radial circulations.  By analyzing the interaction of the particle bed
with the rough pan bottom, we are able to show how particle
entrainment, along with the phase shift relating the horizontal and
vertical vibrations, together conspire to control the direction of the
observed circulations.

Although most previous work on granular circulation has focussed on
either purely vertical or purely horizontal vibrations (see
e.g. Refs.~\cite{JN92,BD92,M94,JNB96,JNBpt96,D98}), a growing body of
research is emerging related to granular flows generated by
simultaneous horizontal and vertical vibrations and/or
shear~\cite{TB98,king00,DB05,got05}.
This is significant since most industrial devices (e.g. seives,
grinders and vibratory conveyors) generate both vertical and
horizontal excitations.  Understanding how the fundamental physical
mechanisms of granular flow relate to practical applications therefore
depends on the detailed examination of systems, such as the one
studied here, having excitations in more than one dimension.  As in
our findings, these earlier works on simultaneous vertical and
horizontal vibrations also point out the significance of the phase
shift between the two types of excitation as being an important
control parameter.

The cylindical geometry of our system (with the cylinder axis oriented
vertically) is also a distinguishing aspect of the present work.  A
few recent works have studied granular flow in annular troughs, and
have also observed tangential circulations similar to those studied
here~\cite{gro04,elhor05}.  However, Ref.~\cite{sistla02} and the
present study are, to our knowledge, the first works to address radial
circulation in this geometry.

\section{\label{sec:pan_model} Description of the pan motion}

\begin{figure}[t]
\includegraphics{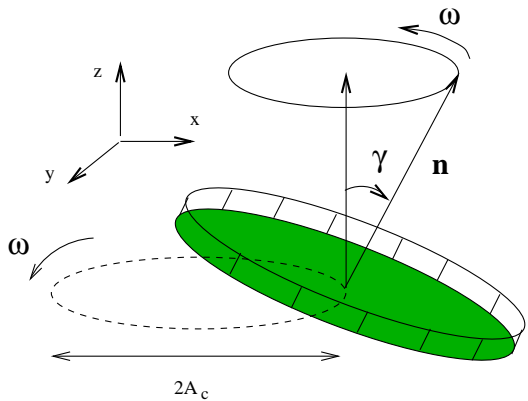}
\caption{Geometry and parameters describing the pan motion. All
  symbols are defined in the text.}
\label{fig:pan_model}
\end{figure}

The shape of the flat-bottomed cylindrical pan and its motion are
illustrated in Fig.~\ref{fig:pan_model}.  The forcing motion we impose
on the pan in our simulations is based on the motion observed in the
experimental apparatus used in Ref.~\cite{sistla02}.  This motion is
complex and is described in detail in this section.

We assume that the pan is a rigid body. Let ${\bf n}$ be the normal
vector to the bottom of the pan, rooted at the center of the bottom
surface.  The motion of the pan can then be described as a
superposition of two motions: (i) a precession of ${\bf n}$ about the
vertical axis with angular frequency $\omega$, and with ${\bf n}$
tilted at a constant angle $\gamma$ with respect to the vertical; and
(ii), a circular motion of the center of the pan in the horizontal
plane, of amplitude $A_c$ and also at frequency $\omega$.  Note that
the direction of rotation of both motions is always the same, and in
our simulations, chosen to be counter-clockwise.

In this motion, the time evolution of the point $(x_c,y_c,z_c)$ at the
center of the pan bottom has only horizontal components which are
non-zero, and thus can be described in the lab frame by,
\begin{eqnarray}
z_c &=& 0 \nonumber \\
x_c &=& A_c\, \cos(\omega t) \label{eq:r_c} \\
y_c &=& A_c\, \sin(\omega t) \nonumber
\end{eqnarray}
Also in the lab frame, the components of the vector ${\bf n}$ are
given by,
\begin{eqnarray}
n_x &=& \sin(\gamma)\, \cos(\omega t - \alpha_{xz} ) \nonumber \\
n_y &=& \sin(\gamma)\, \sin(\omega t - \alpha_{xz} ) \label{eq:n} \\
n_z &=& \cos(\gamma) \nonumber
\end{eqnarray}
where $\alpha_{xz}$ is the phase shift between the circular motion of
the pan center, and the precessional motion of ${\bf n}$.

In order to analyze the behavior of the granular particles in the
moving pan, we also need to derive expressions describing the motion,
in the lab frame, of an arbitrary point $P$ fixed on the pan bottom.
Let $(x_p,y_p)$ be the Cartesian coordinates of $P$ in a frame of
reference fixed to, and lying in the plane of, the pan bottom.  The
coordinates of $P$ in the lab frame, $(x,y,z)$, are then given by,
\begin{eqnarray}
x(t) &=& x_p+A_c \cos(\omega t) \label{eq:xy1} \\
y(t) &=& y_p+A_c \sin(\omega t) \label{eq:xy2} \\
z(t) &=& -\frac{x_pn_x+y_pn_y}{n_z} \label{eq:xy3}
\end{eqnarray}
The expression for $z(t)$ above is determined from the condition that
the pan is a rigid body, and from the orthogonality of the vectors
${\bf n}$ and ${\bf p}$, where ${\bf p}$ is the vector from the center
of the pan bottom to the point $P$, in the pan frame.  If we express
$(x_p,y_p)$ in terms of a polar coordinate system $(r,\theta)$ that is
also fixed in the pan frame, then $x_p=r\cos \theta$ and
$y_p=r\sin\theta$, where $r=\sqrt{x_p^2+y_p^2}$. Using the
trigonometric identity $\cos \alpha \cos \beta + \sin \alpha \sin
\beta = \cos (\alpha - \beta)$, Eq.~\ref{eq:xy3} can be rewritten as,
\begin{equation}
z(t)=-r\,\tan (\gamma)\,\cos (\omega t - \alpha_{xz} -\theta )
\label{eq:z}
\end{equation}

Note that in Ref.~\cite{sistla02}, $A_c$ is referred to as
$x_{max}=y_{max}$ and the maximum vertical amplitude of vibration
$R\sin\gamma$, where $R$ is the radius of the pan, is referred to as
$z_{max}$.  With these changes of notation, the representation of the
pan motion presented in Eqs.~\ref{eq:xy1}, \ref{eq:xy2} and \ref{eq:z}
is consistent with Eqs.~1-3 of reference \cite{sistla02}.

\section{\label{sec:results} Numerical Simulations}

\subsection{\label{sec:ed} Event Driven Algorithm}

The event driven algorithm~\cite{lubachevsky91} used in our
simulations has proved very effective in simulating agitated granular
matter. For example, simulations of standing wave patterns in 3D
vertically oscillated granular layers~\cite{bizon98} show remarkable
agreement with experiments.  The algorithm itself is well described
elsewhere (see e.g. Refs.~\cite{lubachevsky91,lubachevsky92}). In
short, it involves five steps: (i) the calculation of collisions times
between particles and between particles and walls; (ii) determining
which pair of objects are next to collide by finding the minimum of
the collision times; (iii) evolving the system forward in time to the
next-to-occur collision by changing the coordinates and velocities of
all particles according to the equations of motion of free particles
in a gravitational field; (iv) changing the velocities of the two
colliding particles (or of a particle hitting a wall) according to the
conservation of linear and angular momentum and a specified rule for
the dissipation of energy in the collision; and (v), returning to step
(i). The algorithm assumes that interactions occur only through
instantaneous binary collisions.

We have adapted this algorithm to our specific system by implementing
the interactions between particles and moving walls of a cylindrical
flat-bottomed container in a manner that is consistent with hard
sphere dynamics. As described below, we give special care to the
implementation of roughness of the bottom surface of the container. We
have tested several models for the interaction between particles and a
rough bottom plate and decide on the one presented below because of
its simplicity combined with its efficiency.

\subsection{\label{sec:par_space} System Parameters}

In all our simulations the particles are monodisperse spheres of
diameter $\sigma$ possessing both linear and angular momentum.
Otherwise, our numerical model has a fairly large number of parameters
that have to be set so that the model can be validated against the
experimental results of Ref.~\cite{sistla02}.  In the following we
organize these into three groups: dissipative, system size, and
forcing parameters.

{\em Dissipative Parameters:} These parameters control the way energy
is dissipated in the system. We implement inelastic collisions between
particles in the manner described in~\cite{walton93}. The collisions
between particles are thus dependent on the coefficient of restitution
$e$, defined as the ratio of the normal components of the relative
velocities of two colliding particles before and after the collision;
the rotational coefficient of restitution $\beta$, defined as the
ratio of tangential components of relative surface velocities of
colliding particles before and after the collision; and the
coefficient of friction $\mu$ for the material from which the
particles are made.

We set $e$ to be velocity dependent in the following way:
\begin{equation}
e(v_n) = \left\{ \begin{array}{ll}
1-Bv_n^{3/4}, & v_n<v_o \\
\epsilon,       & v_n>v_o
\end{array}
\right.
\label{eq:restitution}
\end{equation}
Here $v_n$ is the component of relative velocity along the line
joining particle centers, $B=(1-\epsilon )v_o^{-3/4},
v_o=\sqrt{g\sigma}$ is a characteristic velocity of the particles, $g$
is the acceleration due to gravity, and $\epsilon$ is a tunable
parameter of the simulations. Decreasing $\epsilon$ increases the
energy dissipated during each collision. The advantages of this model
for the coefficient of restitution are discussed and demonstrated in
Refs.~\cite{shattuck97,bizon98,bizon98b,goldman98,bizon99b,BK05}. 
We have set $\epsilon=0.6$, because this value gives the best
agreement in terms of particle compaction observed in similarly
agitated states in experiments with nylon balls and in our
simulations.

The dissipation of rotational kinetic energy due to colliding
particles is controlled by $\beta$.  As validated in
Ref.~\cite{walton93}, the dependence of $\beta$ on the relative
particle velocities and on $\mu$ depends on the value of the threshold
parameter $\beta_o$ that controls the choice between a rolling or
sliding solution in a given particle-particle
collisions. Ref.~\cite{walton93} suggests that $\beta_o=0.35$ is a
reasonable choice for plastic sphere collisions based on comparison
with experiments. We use this value in all our simulations. We have
set $\mu=0.5$ in all our simulations based on the findings of
Ref.~\cite{bizon98}.

The dissipation of energy in collisions between particles and the side
walls and the bottom of the container is determined by the same set of
parameters ($\epsilon$, $\mu$, and $\beta_o$) as in particle-particle
collisions. For the case of collisions between particles and the side
walls we have observed, however, that the simulated system's behavior
is not detectably changed when using $\epsilon=1$ (elastic collisions)
instead of $\epsilon<1$ (inelastic collisions). Thus, in all our
simulations we set $\epsilon=1$ for collisions between particles and
the side walls.

Special attention is given to the implementation of momentum transfer
and dissipation of energy in collisions between particles and the
rough bottom of the container. Instead of using three dissipative
parameters as in particle-particle collisions, we implement a
one-parameter model as follows.  In the reference frame of the moving
pan, the particle strikes the bottom plate with a relative incident
velocity $\bf{u}$ at the contact point where the surface normal is
$\bf{n}$.  The particle is reflected from the surface with relative
velocity ${\bf v} = e(|{\bf u}|)[{\bf u} - ( 2 {\bf u \cdot n} ) {\bf
  n}] $ where $e(|{\bf u}|)$ has the same form as
Eq.~\ref{eq:restitution}, except that it is a function of the absolute
value of the incident velocity and not just the normal component.
Thus, if not for the factor of $e(|{\bf u}|)$, the particle-bottom
collision would be elastic.  The decrease of all three components of
the reflected velocity by the factor $e(|{\bf u}|)$ is essential here
for modeling the entrainment of particles by the rough bottom
surface. Indeed, the effect of this decrease is that the particle's
velocity (in the lab frame) after the collision tends to be more
similar to the velocity of the rough bottom surface at the point of
contact, than before the collision.  This, in turn, makes the choice
of the reflection direction of little importance.  Statistically the
average direction of reflection must be in the plane of the vectors
$\bf{u}$ and $\bf{n}$, as it is in our model. Thus the only parameter
of our model for the rough bottom surface is $\epsilon$ in
Eq.~\ref{eq:restitution}. We have tested several values of this
parameter in the range between $0.6$ and $0.9$. Lower values of
$\epsilon$ provide better entrainment of particles by the bottom
surface, but at the same time make the average collision time close to
the surface so small that roundoff errors render the simulations
unphysical.  Thus, we have adopted a value of $\epsilon=0.8$ as a
compromise that realizes particle entrainment at the bottom surface as
well as well-behaved numerical simulations.


{\em System Size Parameters:} These parameters include the particle
diameter $\sigma$, number of particles $N$ and the radius of pan
$R$. The layer depth $H$ can be estimated assuming the volume fraction
$\Phi \equiv (N \frac{1}{6}\pi \sigma^3)/(\pi R^2H )$ is equal to the
experimentally determined~\cite{bizon98} value of $0.58$.
For example, in the laboratory experiments of Ref.~\cite{sistla02}
using nylon balls, a 30 kg load of particles (each weighing 0.9~g and
11.5~mm in diameter) in a pan of radius 0.381~m gave $H/\sigma =8.7$
for the layer depth. In the present simulations we set the number of
particles to either $N=2175$ or $N=15000$. In the first case we have
$R/\sigma=12.5$ and $H/\sigma =4.0$, and in the second case we have
$R/\sigma=33$ and also $H/\sigma =4.0$. Although the layer depth in our
simulations is smaller than in the experiments, as we will show below,
we are able to reproduce the main features of the dynamical behavior
observed in experiments. In this work we have chosen to present the
results for only the smaller system size ($N=2175$) because the
shorter computing times allow for larger statistical sampling.  Our
tests of the system behavior using the larger number of particles
indicates that the results are qualitatively the same as for the
smaller system.

{\em Forcing Parameters:} The frequency of vibrations $f=\omega/2\pi$,
the phase shift $\alpha_{xz}$, the amplitude of horizontal vibrations
$A_c$, and the tilt angle $\gamma$, were introduced in Section
\ref{sec:pan_model}. For all results presented here we consider the
set $(f,\alpha_{xz})$ as the independent variables and we study the
system behavior as a function of this set, with all other parameters
fixed. Table~\ref{tab:parameters} shows the values (or range of
values) of the forcing and system size parameters studied here,
compared to those examined in the experiments of Ref.~\cite{sistla02}.

As stated above, we conduct simulations on a system that is smaller
than that of the experiments in order to allow us to study as wide a
range of dynamical states as possible.  Since the pan radius $R$ in
our simulations is therefore smaller than in the experiments, we have
set the value of the tilt angle $\gamma$ to a larger value, such that
the maximum amplitude of vertical motion at the edge of the pan
$z_{max}=R\,\, \tan\gamma$ is approximately the same as in the
experiments.  We also find that we must set the value of $A_c$, the
amplitude of the horizontal motion, to a larger value (compared to
experiments) in order to achieve a robust entrainment of the particles
near the rough bottom surface.  This suggests that the model of
surface roughness used in our simulations is less effective in
entraining particle motion than the real system.

\begin{table}[t]
  \caption{Comparison of simulation parameters used in the present work, 
    and experimental parameters from Ref.~\protect{\cite{sistla02}.}
\label{tab:parameters}}
\begin{ruledtabular}
\begin{tabular}{|c|c|c|c|c|c|c|}
    & $f$ (Hz) & $\alpha_{xz}$ & $A_c/\sigma $  & $\gamma$ & $R/\sigma$ & $H/\sigma$  \\ \hline
exp & 10 - 20   & 0$^\circ$ - 150$^\circ$       & 0.09 - 0.17    & 0.15$^\circ$ - 0.60$^\circ$  & 33.1        & 8.7 \\ \hline
sim &  5 - 55   & 0$^\circ$ - 360$^\circ$        & 0.5         & 1.5$^\circ$  & 12.5      & 4.0
\end{tabular}
\end{ruledtabular}
\end{table}

\section{\label{sec:results1} Results}

\subsection{\label{sec:results-first} Dynamical modes}

\begin{figure}[t]
\includegraphics{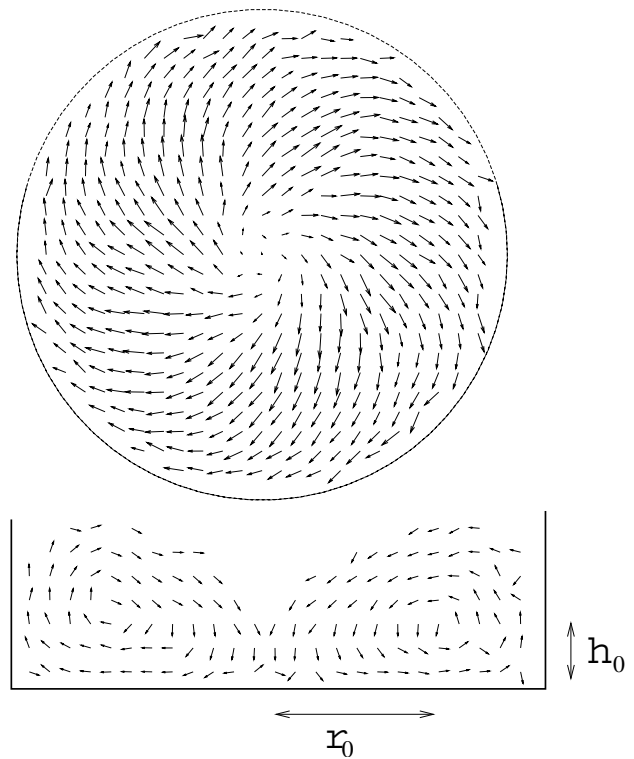}
\caption{\label{fig:vfield_01} Velocity field for toroidal motion at
  $f=25.5$~Hz and $\alpha_{xz} = 144^{\circ}$. Top panel: the
  velocities of particles in the layer nearest the bottom surface,
  viewed from above. Each arrow length is proportional to the
  magnitude of the average velocity corresponding to each grid
  location. Note that the pan center always rotates in the
  counter-clockwise direction in our simulations, but that under the
  conditions of the state shown here, the tangential flow is
  clockwise. Bottom panel: the velocity field of particles of the same
  system in a vertical cross section passing through the pan center.
  Here, all arrow lengths are fixed to a constant, and do not
  represent the magnitude of the average velocities. $r_o$ and $h_o$
  approximately locate the center of circulation of the radial flow.}
\end{figure}

Using the set of parameters described in the previous section, we
study the behavior of the system as a function of $f$ and
$\alpha_{xz}$.  We conduct simulations for 15 different frequencies,
in the range between 5 and 55 Hz. For each $f$ we set the phase angle
$\alpha_{xz}$ to at least 20 different values in the range
$0-360^\circ$. We thereby analyze a total of more then 300 state
points. In the range of $f$ considered here, the motion settles into a
steady state in less than 1.0~s of physical simulation time, which
corresponds to approximately 10-50 revolutions of the pan, depending
on $f$.  To ensure equilibration, we therefore run each simulation for
at least 2~s before starting to accumulate averages.

From a visual examination of animations of the steady-state behavior,
we find that at fixed $\alpha_{xz}$, three main regimes of dynamical
behavior occur as $f$ increases: (i) At small $f$, we observe a
solid-like heap state that undergoes a collective tangential rotation
about the vertical axis.  This is qualitatively the same behavior
observed at small $f$ in Ref.~\cite{sistla02}.  (ii) At larger $f$,
typically in the range of 15-20~Hz, a transition occurs to a
``toroidal'' motion, having simultaneous tangential and radial
circulations, which is also very similar to that observed
experimentally.  (iii) At large $f$, the toroidal state crosses over
to a gas-like state in which the particles have a sufficiently large
kinetic energy that gravity cannot maintain the particles in a dense
bed at the bottom of the pan; in this regime, the particles spread out
over the entire system volume.  This crossover occurs in the range of
$f$ between 30 and 48~Hz, depending on $\alpha_{xz}$.

The simultaneous tangential and radial circulations occurring in the
toroidal regime are shown in Fig.~\ref{fig:vfield_01}.  The top panel
of Fig.~\ref{fig:vfield_01} shows the velocity field for the particles
in a horizontal layer that is adjacent to the pan bottom. The bottom
panel shows the velocity field for particles in a vertical plane
passing through the center of the pan. In these vector field plots we
display the velocity vectors within the corresponding volume of space
averaged over a local volume of $\sigma^3$ and over 400 snapshots of
the system taken 0.001~s apart.  Therefore, the total averaging time
(0.4~s) is much longer than the period of oscillation of the pan in
the range of $f$ studied here.

\subsection{\label{sec:phd_2.175K}Phase diagram}

\begin{figure}[!ht]
\includegraphics[width=2.6in]{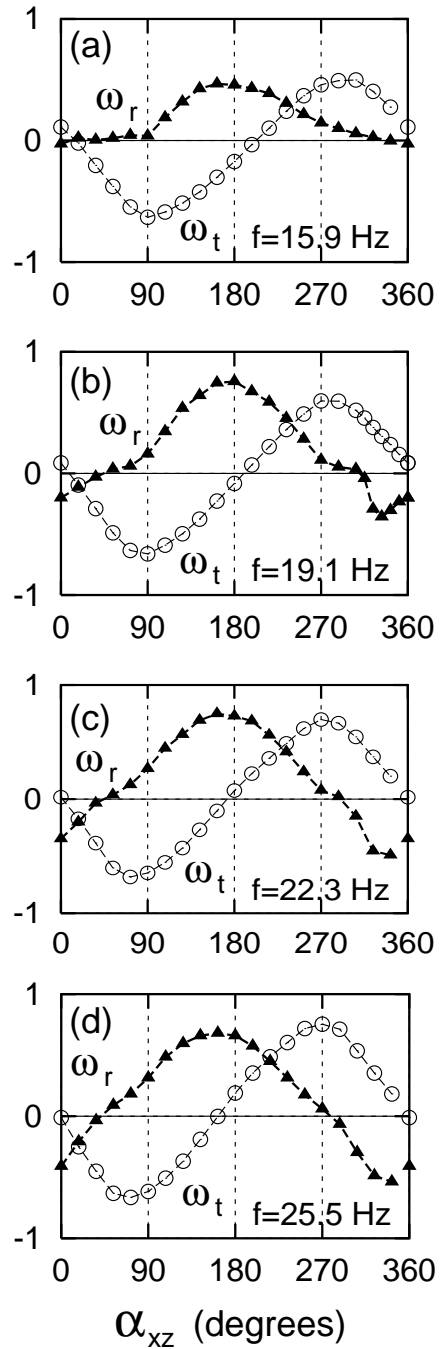}
\caption{\label{fig:plot_w} $\omega_r$ (triangles) and $\omega_t$
  (circles) as a function of $\alpha_{xz}$ at various $f$.  To
  facilitate comparison, $\omega_r$ values are multiplied by 30, and
  $\omega_t$ values by 1.5.}
\end{figure}

\begin{figure}[t]
\includegraphics{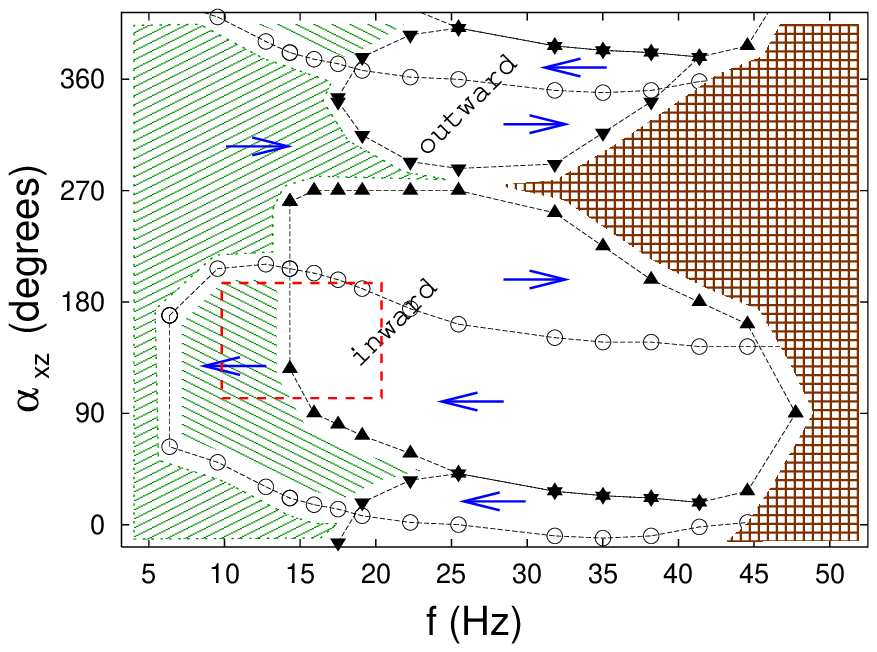}
\caption{\label{fig:phase_d_N2.175K} Phase diagram of dynamical modes
  found in simulations. The solid-like heap state occurs at low $f$,
  in the diagonally striped areas. The domain of the gas-like state at
  high $f$ is denoted by the square lattice pattern.  The remaining
  white areas, bounded by triangles, indicate the domains in which the
  toroidal circulation is observed.  The label ``inward'' indicates
  the region where the radial flow is such that particles travel
  toward the center along the top surface of the bed.  The label
  ``outward'' indicates the region where the radial flow is such that
  particles travel away from the center along the top surface of the
  bed.  For both the heap and toroidal states, the line that connects
  the open circles indicates a change of direction of the tangential
  flow. Left (right) arrows indicate that the direction of the
  tangential flow is against (with) the direction of the pan
  rotation. The dashed rectangle shows the approximate region
  corresponding to the experiments of
  Ref.~\protect{\cite{sistla02}}. Error bars for the boundaries of the
  toroidal domains (triangles) are of the size of the symbol for low
  $f$ and increase to up to three times the size of the symbol for
  high $f$.}
\end{figure}

\begin{figure}[t]
\includegraphics{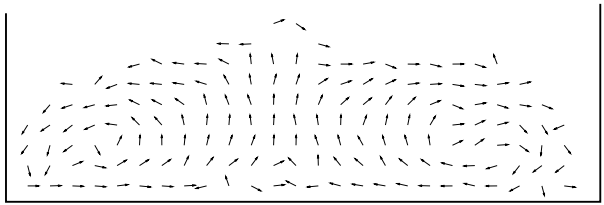}
\caption{\label{fig:vfield_02} Velocity field in a vertical cross
  section passing through the pan center, illustrating {\it outward}
  radial flow, occurring at $f=27$~Hz and $\alpha_{xz}=0^\circ$. }
\end{figure}

We next seek to create a dynamical phase diagram of the system,
i.e. to subdivide the plane of $f$ and $\alpha_{xz}$ into domains
according to where each of the observed dynamical modes occurs.  To do
so, we quantify the state of circulation in the granular bed using two
measures of the average velocity of the particles.

The first measure, $\omega_t$, quantifies the tangential circulation
and is the average angular velocity of particles in the horizontal
plane around the vertical line of symmetry:
\begin{eqnarray}
  \omega_t = \frac{1}{N} \sum_i \frac{v_{t,i}}{d_i} =
  \frac{1}{N} \sum_i \frac{1}{d_i} \,
  {\bf{v}}_i \cdot (\hat{\bf{k}} \times \hat{\bf{r}}_i )
\label{eq:w_l}
\end{eqnarray}
Here, $v_{t,i}$ is the tangential component of ${\bf v}_i$, the
velocity of particle $i$, about the vertical axis.  $d_i$ is the
perpendicular distance from a particle $i$ to the $z$-axis of the lab
frame; $\hat{\bf{k}}$ is the unit vector pointing along the $z$-axis
of the lab frame; and $\hat{\bf{r}}_i$ is the radial unit vector at
the position of particle $i$ in the lab frame.

The second measure quantifies the radial circulation.  We define this
measure, $\omega_r$, as the average angular velocity of particles
about the curved axis that forms a horizontal circle of radius $r_o$
and elevated above the bottom of the pan at a distance $h_o$.  As
illustrated in the lower panel of Fig.~\ref{fig:vfield_01}, if we
choose $r_o=0.75R$ and $h_o=0.75H$, we find that this curved axis
approximately coincides with the center of the radial circulation
observed in our simulations.  $\omega_r$ is then given by,
\begin{eqnarray}
  \omega_r = \frac{1}{N} \sum_i \frac{v_{r,i}}{a_i} =
  \frac{1}{N} \sum_i \frac{1}{a_i} \,
  {\bf{v}}_i \cdot (\hat{\bf{t}}_i \times \hat{\bf{a}}_i )
\label{eq:w_o}
\end{eqnarray}
Here $v_{r,i}$ is the tangential component of ${\bf v}_i$, in a
vertical plane, with respect the circulation center defined by $r_o$
and $h_o$; and $\hat{\bf{t}}_i = (\hat{\bf{k}} \times
\hat{\bf{r}}_i)$.  $\hat{\bf{a}}_i$ and $a_i$ are respectively the
direction and magnitude of the vector ${\bf{a}}_i = {\bf{r}}_i - (h_o
\hat{\bf{k}} + r_o \hat{\bf{r}}_i ) $ which defines the position of
particle $i$ with respect to the center of circulation in the vertical
plane.

Examples of the behavior of $\omega_t$ and $\omega_r$, as a function
of $\alpha_{xz}$ at various fixed $f$, are shown in
Fig.~\ref{fig:plot_w}.  In Fig.~\ref{fig:plot_w}(a), we see that
$\omega_r=0$ for $\alpha_{xz}<90^\circ$, becomes non-zero for higher
$\alpha_{xz}$, but then returns to zero as $\alpha_{xz}$ approaches
$360^\circ$.  Based on our visual observations of the computer
animations, the heap state exhibits no radial circulation, and so the
progression of behavior of $\omega_r$ in Fig.~\ref{fig:plot_w}(a)
reflects the transition first from the heap to the toroidal mode, and
then back again, as a function of $\alpha_{xz}$.  On the other hand,
we see that $\omega_t$ varies smoothly as a function of $\alpha_{xz}$
in Fig.~\ref{fig:plot_w}(a), reflecting the fact that tangential
circulation is common to both the heap and toroidal modes.  Based on
these observations, for the purpose of constructing a phase diagram,
we define the toroidal state as that in which $\omega_r$ is clearly
different from zero.  We also note that the variation of $\omega_t$
and $\omega_r$ with $\alpha_{xz}$ is approximately sinusoidal, with
the behavior of $\omega_r$ lagging that of $\omega_t$ by about
$90^\circ$.  We will return to this behavior in the next section.

The resulting phase diagram is shown in
Fig.~\ref{fig:phase_d_N2.175K}.  The most notable behavior of the
observations summarized in Figs.~\ref{fig:plot_w} and
\ref{fig:phase_d_N2.175K} is that we find that both the radial and
tangential circulations reverse their direction as a function
(primarily) of $\alpha_{xz}$, as indicated by the occurrence of both
positive and negative values of $\omega_t$ and $\omega_r$ in
Fig.~\ref{fig:plot_w}. The result is a rich and highly controllable
spectrum of dynamical behavior: at low $f$, we see heap states that
rotate either clockwise or counter-clockwise (viewed from above); at
higher $f$, toroidal states with any combination of clockwise or
counter-clockwise tangential circulation, and inward or outward radial
circulation (based on the flow direction on the upper free surface)
can be realized.  The velocity field of a system with an outward
radial circulation is shown in Fig.~\ref{fig:vfield_02}.  In addition,
the magnitudes of both $\omega_t$ and $\omega_r$ can be readily tuned
through their dependence on $f$ and $\alpha_{xz}$.

The phase behavior shown in Fig.~\ref{fig:phase_d_N2.175K} is in
quantitative agreement with the experimental results of
Ref.~\cite{sistla02} for the boundary between the heap and toriodal
modes.  The forcing conditions studied in Ref.~\cite{sistla02}
correspond approximately to $10<f<20$ and
$100^\circ<\alpha_{xz}<190^\circ$, in the notation of this work.  In
this range, the tangential circulation is (almost always) in the same
direction for both the heap and toroidal modes, and, as in
experiments, is opposite to the direction the pan center rotates.  The
exception, in both simulations and experiments, occurs for small
values of both $f$ and $\alpha_{xz}$, where the tangential circulation
occurs in the same direction as the pan center rotates.  Since the
experiments only studied a small fraction of the phase diagram
explored here, the reversal of the radial circulation was not observed
in experiments, but is revealed here, along with the complete
demarcation of the reversal boundary for the tangential circulation.

We note that the more subtle modifications of the toroidal mode
observed experimentally (the ``surface waves'' and ``sectors''
described in Ref.~\cite{sistla02}) are not observed here, suggesting
that these phenomena depend on aspects of the motion and/or particle
interactions that are not modeled in the present simulations.  One
possibility is that these modifications of the toroidal mode are due
to variations of the pan motion from a perfect sinusoid.  As indicated
in Ref.~\cite{sistla02}, the experimental apparatus was not perfectly
symmetrical, and the measured motion indicated the presence of a weak
pattern of beats superimposed on the principal motion.  While it is
possible that the experimentally observed surface waves and sectors
had there origin in these additional components of the motion, more
research, beyond the scope of that described here, would be required
to clarify this issue.

\section{Origins of observed flows}

In this section, we discuss some of the physical mechanisms that may
underlie the behavior noted so far.  We focus in turn on two
questions: (i) Why do the horizontal and radial circulations reverse
their direction as a function of $\alpha_{xz}$? (ii) What parameters
determine the onset of heap-to-toroidal flow?

\subsection{Circulation reversals}

\begin{figure}[t]
\includegraphics{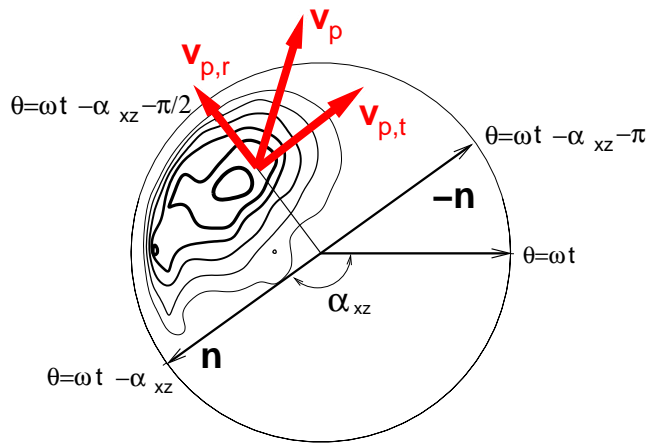}
\caption{\label{fig:pan-model-top} Interrelationship of the zone of
  contact between the granular bed and the pan bottom, and the motion
  of the pan, as viewed from above, for $f=23.4$~Hz,
  $\alpha_{xz}=144^\circ$.  The contours are lines along which the
  average momentum change of particles colliding with the bottom
  surface is constant.  The vector labeled ${\bf n}$ is the
  projection of the pan normal vector into the horizontal plane, and
  indicates the direction along which the vertical displacement of the
  pan is most negative.  The vector labelled ${\bf -n}$ therefore
  indicates the direction along which the vertical displacement of the
  pan is most positive.  Since the pan center orbits counter-clockwise
  in the horizontal plane, the half circle above and to the left of
  the bisector formed by ${\bf n}$ and ${\bf -n}$ is that half of the
  bottom surface that is rising.  The zone of contact, located in the
  region around the center of the momentum-change contours, is located
  on the rising side of the pan bottom, and has its center near the
  direction $\theta_o = \omega t -\alpha_{xz}- \pi/2$.  The horizontal
  ``pushing velocity'' of the bottom pan surface, ${\bf v}_p$, and its
  polar component vectors, are also shown.  }
\end{figure}

In order to clarify the nature of the interaction between the granular
bed and the moving pan, we first examine the nature of the contact
between the two as a function of time.  To determine where and when
the bed is in contact with the pan as a function of time, we subdivide
the pan bottom into a grid of squares of area $\sigma^2$.  Over a time
interval of 0.05 of the period of revolution of the pan, we then find
the average change of momentum of particles colliding with the pan
bottom in each grid square.  This provides a measure of the
instantaneous pressure exerted by the bed on each area element of the
pan bottom.  When the system is in a dynamical steady state, we can
average over ``snapshots'' of the pressure distribution on the pan
bottom at different times by rotating each snapshot around the pan
center such that the direction given by $\theta=\omega t$ always
points along the positive $x$-axis.  Pressure contours resulting from
this averaging process over a total time of 300 revolutions of the pan
are shown in
Fig.~\ref{fig:pan-model-top} for the case $f=23.4$~Hz,
$\alpha_{xz}=144^\circ$.  The pressure contours show that a
well-defined ``zone of contact'' exists between the granular bed and
the pan bottom at a particular angular position.  Similar results are
obtained at other values of $f$ and $\alpha_{xz}$, and differ only in
the angular position of the zone of contact.

We find that $\theta_o$, the angular position of the center of the
zone of contact is, in all cases examined, close to the value,
\begin{equation}
  \theta_o = \omega t -\alpha_{xz}- \pi/2
\label{eq:theta_c}
\end{equation}
A relatively simple rationale supports this expression: We expect that
the contact between the bed and the pan occurs on that side of the pan
which is rising from its maximum negative displacement in the $z$
direction to its maximum positive displacement.  As shown in
Fig.~\ref{fig:pan-model-top}, when the normal vector of the pan ${\bf
  n}$ is precessing counter-clockwise, the ``rising'' half of the pan
bottom is identified by those values of $\theta$ such that,
\begin{equation}
0 < (\omega t -\alpha_{xz}-\theta ) < \pi
\label{eq:c_range}
\end{equation}
We see from Fig.~\ref{fig:pan-model-top} that the maximum pressure of
the zone of contact is approximately in the middle of this half of the
pan bottom, the angular position of which is given approximately by
Eq.~\ref{eq:theta_c}.

Next we note that we have modeled the pan bottom as a rough surface
that entrains the motion of particles that collide with it; i.e.
particles tend to have a post-collision velocity with a direction
similar to that of the surface element of the pan bottom that they
strike.  Since the contact between the granular bed and the pan bottom
is localized to the zone of contact, we expect that the velocity of
the surface elements at the angular position given by
Eq.~\ref{eq:theta_c} will have a significant influence on the
direction of flow of the bed.  Let us define the ``pushing velocity''
${\bf v}_p$ as the component of the pan bottom velocity in the
horizontal plane at $\theta=\theta_o$.  The time derivatives of
Eqs.~\ref{eq:xy1} and \ref{eq:xy2} give us the $x$ and $y$ components
of ${\bf v}_p$: $v_{p,x}=-A_c\omega \sin(\omega t)$ and
$v_{p,y}=A_c\omega \cos(\omega t)$.  Converting these Cartesian
components to those in a polar coordinate system gives the radial and
tangential components of ${\bf v}_p$.  The desired radial unit vector
is $\hat {\bf r} = (\cos\theta_o, \sin\theta_o)$ and the tangential
unit vector is $\hat {\bf t} = [\cos(\theta_o+\pi/2),
\sin(\theta_o+\pi/2)]$.  The radial and tangential components of ${\bf
  v}_p$ are therefore given by,
\begin{eqnarray}
v_{p,r} = {\bf v}_p{\bf \cdot \hat r} = -A_c \omega \cos (\alpha_{xz})
\label{eq:theory_overs} \\ 
v_{p,t} = {\bf v}_p{\bf \cdot \hat t} = -A_c \omega \sin (\alpha_{xz}) 
\label{eq:theory_laps}
\end{eqnarray}

Comparing these expressions with the dependence of $\omega_t$ and
$\omega_r$ on $\alpha_{xz}$ shown in Fig.~\ref{fig:plot_w} shows that
there is a good correspondence between the polar components of ${\bf
  v}_p$ and the observed angular velocity of both the radial and
tangential flows.  $\omega_t$ shows an especially good match to a
function proportional to $-\sin(\alpha_{xz})$.  The correspondence of
$\omega_r$ to $-\cos(\alpha_{xz})$ is complicated by the fact that
$\omega_r=0$ in the heap state.  However the observed trend for
$\omega_r$ to lag $\omega_t$ by $90^\circ$ is consistent with the
implications of Eqs.~\ref{eq:theory_overs} and \ref{eq:theory_laps}.

Overall, we therefore find that the assumption that the particles are
entrained by the pan bottom in a specific zone of contact explains
well the observed circulation.  If the polar components of ${\bf v}_p$
drive the motion observed via $\omega_t$ and $\omega_r$, then
Eqs.~\ref{eq:theory_overs} and \ref{eq:theory_laps} predict the
occurrence of both clockwise and counter-clockwise tangential
circulations, as well as both inward and outward radial flows.  In
particular, Eqs.~\ref{eq:theory_overs} and \ref{eq:theory_laps} also
provide a basis for understanding why the tangential flow can be both
the same as, or opposite to, the direction of rotation of the pan
center.  Most importantly, the dominant role of the phase shift
$\alpha_{xz}$ in controlling the direction of the observed flows is
confirmed.

\subsection{Onset of toroidal flow}

\begin{figure}[t]
\includegraphics{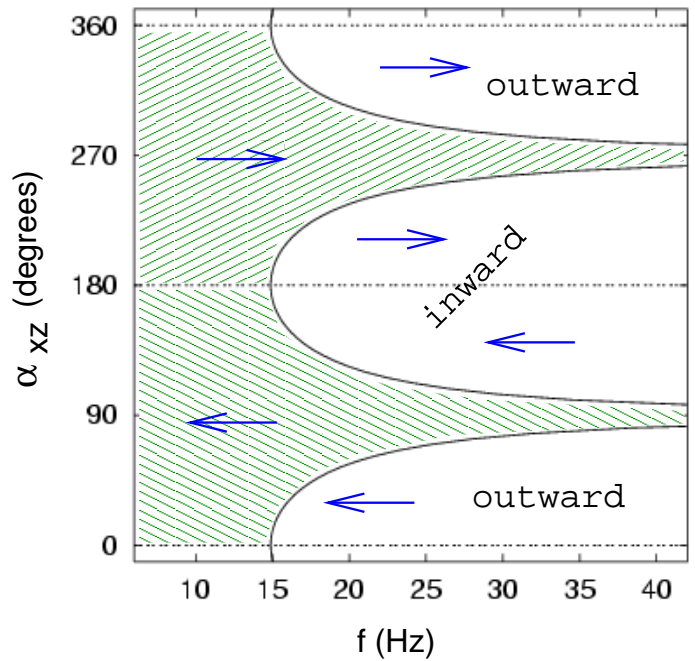}
\caption{\label{fig:phd_theory} Model phase diagram based on
  Eqs.~\ref{eq:theory_overs}, \ref{eq:theory_laps} and \ref{eq:wc}.
  Solid lines indicate the onset of radial flow. Dotted lines indicate
  the change of direction of the tangential flow.  Arrows and region
  labels have the same meaning as in
  Fig.~\protect{\ref{fig:phase_d_N2.175K}}.}
\end{figure}

Since both the heap and toroidal circulations exhibit tangential flow,
understanding the onset of the toroidal mode requires us to understand
the conditions required for the appearance of the radial circulation.
This is a complex question, since the heap-to-toroidal boundary in
Fig.~\ref{fig:phase_d_N2.175K} is clearly dependent both on
$\alpha_{xz}$ and $f$.  

To make a preliminary attempt to understand the dependence of the
heap-to-toroidal boundary on $\alpha_{xz}$ and $f$, let us consider
$a_r$, the maximum radial acceleration experienced by the particle bed,
which drives the radial circulation observed in the toroidal mode.  By
analogy with many other systems exhibiting granular circulation, we
expect that the onset of the toroidal mode occurs when $a_r$ exceeds a
critical threshold large enough to sustain the radial circulation.

To estimate the scaling of $a_r$ with $\alpha_{xz}$ and $f$, we first
assume that $a_r=A_p\omega^2$, where $A_p$ is the average displacement
of the particle bed in the radial direction as it is pushed by the
rough pan bottom. Given the correspondence between the components of
the pushing velocity ${\bf v}_p$ and the circulation speeds $\omega_t$
and $\omega_r$ established in the previous section, we propose that
$A_p$ is proportional to the product of $|v_{p,r}|$ and $\tau_p$, the
time interval during which particles are exposed to the zone of
contact.  We also assume that $\tau_p$ is simply some fraction of the
period of the pan motion, and so is proportional to $2\pi/\omega$.
Combining these assumptions with the expression for $v_{p,r}$ in
Eq.~\ref{eq:theory_overs}, we therefore can write,
\begin{equation}
  a_r \propto \omega^2 |\cos \alpha_{xz}| 
\label{eq:def_Gamma}
\end{equation}

If the onset of the heap-to-toroidal transition occurs at a critical
value of $a_r=a_{r,c}$ then the corresponding critical angular
frequency $\omega_c$ for the onset of radial circulation (and
therefore toroidal flow) is given by,
\begin{equation}
  \omega_c \propto \sqrt{\frac{a_{r,c}}{|\cos \alpha_{xz}|}}
\label{eq:wc}
\end{equation}
Fig.~\ref{fig:phd_theory} shows a plot of the critical frequency
$f_c=\omega_c/2\pi$ in the $f$-$\alpha_{xz}$ plane, based on the
proportionality expressed in Eq.~\ref{eq:wc}, with the constant of
proportionality chosen so that the mimimum onset frequency is $15$~Hz,
as found in our simulations.  Comparing
Figs.~\ref{fig:phase_d_N2.175K} and \ref{fig:phd_theory} shows that
the model expressed in Eq.~\ref{eq:wc} captures several qualitative
features of the observed variation of the heap-to-toroidal transition:
e.g. the occurrence of maxima in the transition boundary as a function
of $\alpha_{xz}$; and the separation of the toroidal regime into two
distinct domains, one exhibiting an inward radial flow, and one
outward.

Clearly the approximations of this simple model should be refined if
it is to have any quantitative predictive value.  However, it does
support the view that entrainment of the particle bed by the rough
bottom is a key factor in determining the overall behavior of the
system.

\section{\label{sec:conclusions} Discussion and Conclusion}

\begin{figure}[t]
\includegraphics{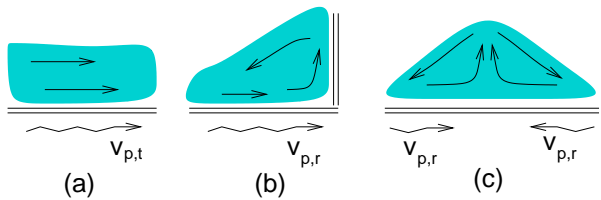}
\caption{\label{fig:boundary3} Three types of boundary condition
  occurring in this system: (a) the periodic boundary condition
  involving the tangential flow; (b) the vertical wall obstacle
  encountered by the inward radial flow; (c) central ``plume''
  obstacle encountered by the outward radial flow.}
\end{figure}

In summary, we have established the following picture for the
dynamical behavior observed in our simulations, and to a large degree,
in the experiments of Ref.~\cite{sistla02}: The granular bed is
excited by a rough horizontal surface that is undergoing both vertical
and horizontal vibrations. The vertical vibrations impose a
periodicity on the contact between the bed and the surface, and the
horizontal vibrations push the granular bed only during the interval
of contact. The phase shift relating the vertical and the horizontal
components of the vibration determines the direction and strength of
the horizontal pushing force applied by the rough surface on the bed.
The result is a highly tunable spectrum of behavior, in which both
tangential and radial circulations, in arbitrary combinations, can be
generated.

The cylindrical geometry studied here plays an important underlying
role in the granular flows we observe.  The tangential circulation is
in many ways similar to the kind of linear flow observed in vibratory
conveyors~\cite{hongler89,sloot96}, but wrapped around into a
circle, creating in effect a periodic boundary condition. Indeed, a
recent work studied the flow of granular matter in an annular
vibratory conveyor and, as in our work, noted both the occurrence of a
reversing tangential flow, and the controlling influence on this flow
of the phase shift between vertical and horizontal
vibrations~\cite{gro04}.  The fact that the granular bed in our system
encounters no obstacles as it moves about the center accounts for the
ease with which the tangential flow occurs not only in the toroidal
mode, but also the heap mode at very small $f$.

In contrast, the granular flow in the radial direction encounters a
vertical barrier that it must surmount: either the outer wall in the
case of the inward radial flow, or the central ``plume'' that is
created when the granular particles are pushed toward the center in
the outward radial flow (see Fig.~\ref{fig:boundary3}).  Here, the
mechanism that drives the flow along the bottom surface is also
similar to that of a vibratory conveyor, but with the complication
that this driving force must be large enough to overcome the
gravitational barrier imposed at the vertical boundary. This largely
explains why the radial flow only occurs above a critical value of
$f$.

Our results also demonstrate that complex dynamical modes observed in
real (i.e. industrial) granular matter devices can, on the one hand,
be understood using computer simulations; and on the other hand, the
behavior observed in these real devices can reveal important
fundamental principles (such as the role of the phase factor
established here) relating to the creation and control of highly
specialized granular flows.

\bigskip
\begin{acknowledgments}
  We acknowledge useful discussions with K. Bevan, S. Fohanno,
  P. Sistla and E.B. Smith.  We thank Materials and Manufacturing
  Ontario and NSERC (Canada) for financial support; and SHARCNET for
  computing resources.  PHP acknowledges the support of the CRC
  Program.
\end{acknowledgments}


\end{document}